\documentclass[aps,twocolumn,tightenlines]{revtex4}
\usepackage[dvipdf]{graphicx}

\begin{document}
\title{First measurements of the index of refraction of gases for lithium atomic waves}

\author{M. Jacquey, M. B\"uchner, G. Tr\'enec and J. Vigu\'e}
\address{ Laboratoire Collisions Agr\'egats R\'eactivit\'e -IRSAMC
\\Universit\'e Paul Sabatier and CNRS UMR 5589
 118, Route de Narbonne 31062 Toulouse Cedex, France
\\ e-mail:~{\tt jacques.vigue@irsamc.ups-tlse.fr}}

\date{\today}

\begin{abstract}

We report the first measurements of the index of refraction of
gases for lithium waves. Using an atom interferometer, we have
measured the real and imaginary parts of the index of refraction
$n$ for argon, krypton and xenon, as a function of the gas density
for several velocities of the lithium beam. The linear dependence
of $(n-1)$ with the gas density is well verified. The total
collision cross-section deduced from the imaginary part of $(n-1)$
is in very good agreement with traditional measurements of this
quantity. Finally, the real and imaginary parts of $(n-1)$ and
their ratio $\rho$ exhibit glory oscillations, in good agreement
with calculations.

\end{abstract}
\maketitle


The concept of the index of refraction for waves transmitted
through matter was extended from light waves to neutron waves
around 1940, as reviewed by M. Lax \cite{lax51}. The extension to
atom waves has been done by D. Pritchard and co-workers, with the
first measurements of the index of refraction of gases for sodium
waves \cite{schmiedmayer95} in 1995 and the subsequent observation
of glory oscillations on the index variations with sodium velocity
\cite{schmiedmayer97,hammond97,roberts02,roberts02a}. We report
here the first measurements of the index of refraction of gases
for lithium waves.

Several papers \cite{forrey96,forrey97,leo00,kharchenko01,
forrey02,blanchard03,vigue95,audouard95,champenois97,champenois99}
have dealt with the theory of the index of refraction $n$. The
index of refraction is proportional to the forward scattering
amplitude, which can be calculated if the interaction potential
between an atom of the wave and an atom of the target gas is
known. The imaginary part of the forward scattering amplitude is
related to the total cross section but its real part can be
measured only by atom interferometry. This amplitude exhibits
resonances, for a collision energy comparable to the potential
well depth, and glory oscillations, for larger energy. These glory
oscillations are due to the existence of a undeflected classical
trajectory resulting from the compensation of attractive and
repulsive forces \cite{pauly79}.

A measurement of the index of refraction thus provides a new
access to atom-atom interaction potentials. Many other experiments
are sensitive to the atom-atom interaction potentials: in the
particular case of alkali-rare gas pairs, measurements of total
and differential cross sections, line broadening experiments and
spectroscopy of van der Waals molecules have been much used. Each
technique is more sensitive to a different part of the potential
curve and one would expect that very accurate potentials are
available, but, as shown by the calculations done by D. Pritchard
and co-workers and by our research group
\cite{roberts02,champenois97,champenois99}, the index of
refraction deduced from various potentials differ substantially,
thus proving the need for more accurate potentials.

Our experiment is similar to the experiment of D. Pritchard and
co-workers \cite{schmiedmayer95,schmiedmayer97,roberts02}. We have
measured separately the real and imaginary parts of $(n-1)$ with a
good accuracy and tested their linear dependence with the gas
density. The total collision cross-section deduced from our
measurement of the imaginary part ${\mathcal{I}}m(n-1)$ is in very
good agreement with previous measurements by L. Wharton and
co-workers \cite{ury72,dehmer72}. Our measurements of the real and
imaginary parts of $(n-1)$ and of their ratio $\rho$ are in good
agreement with the calculations done by C. Champenois
\cite{champenois99}, using potential curves fitted by L. Wharton
and co-workers \cite{ury72,dehmer72}.

\begin{figure}
\includegraphics[width = 7.5 cm,height= 5.6 cm]{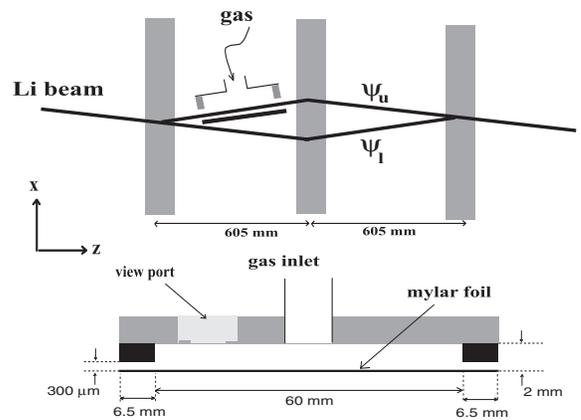}
\caption{Upper part: schematic drawing of a top view of the
interferometer, with the gas cell inserted just ahead of the
second laser standing wave. Lower part: top view of the gas cell.
The view-port is used to align the septum by optical techniques.
The slit widths are exaggerated to be visible.}\label{fig1}
\end{figure}

The principle of the experiment is to introduce some gas on one of
the atomic beams inside an atom interferometer, as represented in
Fig. \ref{fig1}. Noting $\psi_{u/l}$ the waves propagating on the
upper/lower paths inside the interferometer, the interference
signal $I$ is given by:

\begin{equation}
\label{n0} I = \left|\psi_{l} + \psi_{u} \exp\left( i
\varphi\right) \right|^2
\end{equation}

\noindent The phase $\varphi = k_G(2 x_2 - x_1 - x_3)$, which
depends on the grating positions $x_i$ ($k_G$ is the grating
wavevector), is used to observe interference fringes. We can
rewrite equation (\ref{n0}):

\begin{equation}
\label{n1} I = I_B +I_0 \left[ 1 + {\mathcal{V}} \cos
\left(\varphi\right) \right]
\end{equation}

\noindent $I_0$ is the mean intensity, ${\mathcal{V}}$ the fringe
visibility and we have added the detector background $I_B$. When
the atomic wave propagates in a gas of density $n_{gas}$, its wave
vector $\mathbf{k}$ becomes $n \mathbf{k}$, where $n$ is the index
of refraction. For a gas cell of length $L$ in the upper path, the
wave $\psi_{u}$ is replaced by the transmitted wave $\psi_{u,t}$
given by:

\begin{equation}
\label{n2} \psi_{u,t} / \psi_{u} = \exp\left[i \left(n-1\right)k
L\right] = t(n_{gas})\exp\left[i \varphi(n_{gas})\right]
\end{equation}

\noindent with $t(n_{gas}) = \exp\left[-{\mathcal{I}}m(n-1) k
L\right]$ and $\varphi(n_{gas}) = {\mathcal{R}}e(n-1)k L$. The
signal given by equation (\ref{n1}) is modified, with a phase
shift $\varphi(n_{gas})$. The mean intensity $I_0(n_{gas})$ and
the fringe visibility ${\mathcal{V}}(n_{gas})$ are both changed
and $t(n_{gas})$ is related to these quantities by:

\begin{equation}
\label{n3} t(n_{gas}) =
I_0(n_{gas}){\mathcal{V}}(n_{gas})/\left[I_0(0){\mathcal{V}}(0)\right]
\end{equation}

Our Mach-Zehnder atom interferometer uses laser diffraction in the
Bragg regime \cite{miffre05}, with a laser wavelength close to the
lithium first resonance line at $671$ nm. To optimize the signal
for the $^7$Li isotope with first order diffraction, we have used
a frequency detuning equal to $4$ GHz, a total laser power equal
to $275$ mW and a beam waist radius $w_0 = 6.2$ mm. The lithium
beam mean velocity $u$ is varied by seeding lithium in a rare gas
mixture and the velocity distribution thus achieved has a full
width at half maximum close to $25$\%.

\begin{figure}[b]
\includegraphics[width = 8 cm,height= 6 cm]{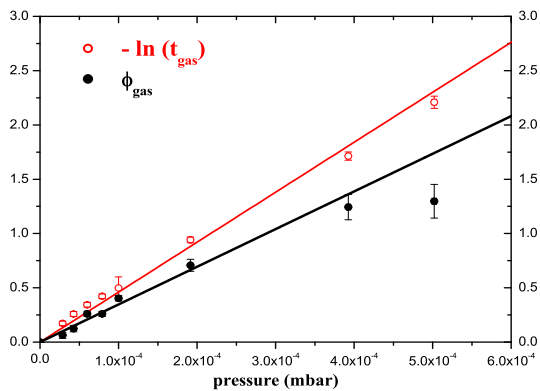}
\caption{Plot of the measured values of the phase shift
$\varphi(n_{gas})$ and of the logarithm of the amplitude
transmission $-\ln\left(t(n_{gas}\right))$ as a function of xenon
pressure $p_{cell}$. The lithium beam mean velocity is $u= 1075
\pm 20 $ m/s.}\label{fig2}
\end{figure}

For a mean atom velocity $u= 1075$ m/s, the maximum distance
between the atomic beam centers, close to $100$ $\mu$m, is
sufficient to insert a septum between these beams. The septum, a
$6$ $\mu$m thick mylar foil, separates the gas cell from the
interferometer vacuum chamber. This cell is connected to the
interferometer chamber by $300$ $\mu$m wide slits as shown in Fig.
\ref{fig1}, in order to reduce the gas flow. The cell is connected
by a $16$ mm diameter ultra-high vacuum gas line to a leak valve
used to introduce the gas. An other valve connects the gas line to
the interferometer vacuum chamber so that the cell pressure can be
reduced to its base value in about $3$ s. We use high purity gases
from Air Liquide (total impurity content below $50$ ppm) and a
cell pressure in the $10^{-3}-10^{-4}$ millibar range. The
pressure is measured by a membrane gauge (Leybold CERAVAC CTR91),
with an accuracy near $1$\%. When the cell is evacuated, the
measured pressure, $p_{meas.} = (1\pm 1)\times 10^{-6}$ millibar,
is negligible. Because of the gas flow through the connection
pipe, the cell pressure $p_{cell}$ is slightly less than the
measured value $p_{meas.}$ as the gauge is at about $50$ cm from
the cell. Molecular flow theory predicts that the pressure in the
cell $p_{cell}$ is homogeneous within $1$\% and that
$p_{cell}/p_{meas.} = C_{pipe}/(C_{pipe} + C_{slits})$.  The
conductances of the pipe, $C_{pipe}$, and of the slits,
$C_{slits}$, have been calculated and we get $p_{cell}/p_{meas.}=
0.90 \pm 0.01$. We then use the ideal gas law at $T = 298$ K to
deduce the gas density $n_{gas}$ in the cell.

We record interference fringes by displacing the third standing
wave mirror with a linear voltage ramp applied on a piezo-electric
stage. In order to correct the interferometer phase drift, each
experiment is made of three sweeps, the first and third ones
($j=1$ and $3$) with an empty cell and the second one ($j=2$) with
a pressure $p_{cell}$. The counting time is $0.3$ s per data
point, with $300$ points per sweep. After the third sweep, we flag
the lithium beam to measure the detector background $I_B$. We
assume that the phase $\varphi$ can be written $\varphi= a_j + b_j
n + c_j n^2$, where the quadratic term describes the non-linearity
of the piezo stage ($n$ being the channel number). The best fit of
each recording, using equation (\ref{n1}), provides the initial
phase $a_j$, the mean intensity $I_{0j}$ and the fringe visibility
${\mathcal{V}}_j$. We thus get the effect of the gas, namely the
phase shift $\varphi(n_{gas}) =a_2 -(a_1 + a_3)/2$, and the
attenuation $t(n_{gas})$ given by equation (\ref{n3}) (the
$I_{0}(0){\mathcal{V}}(0)$ value is taken as the mean of the $j=1$
and $j=3$ values).

\begin{table}[h]
\begin{tabular}{c|ccc}
gas&Ar&Kr&Xe\\
\hline
$10^{29}{\mathcal{R}}e(n-1)/n_{gas}$&$1.20\pm0.11$&$1.57\pm0.10$&$1.82\pm0.07$\\
$10^{29}{\mathcal{I}}m(n-1)/n_{gas} $&$2.11\pm0.06$&$1.99\pm0.07$&$2.40\pm0.06$\\
\hline
$\rho$& $0.56\pm 0.05$& $0.78\pm 0.04$&$0.70\pm0.03$\\
 \hline
\end{tabular}
\caption{\label{index1065} Index of refraction of argon, krypton
and xenon for lithium waves with a mean velocity $u = 1075\pm 20$
m/s. For each gas, we give the real and imaginary parts of
$10^{29}(n-1)/ n_{gas}$ ($n_{gas}$ in m$^{-3}$) and the ratio
$\rho = {\mathcal{R}}e(n-1) /{\mathcal{I}}m(n-1)$.}
\end{table}

For a given lithium mean velocity $u$, we measure the phase shift
and the amplitude attenuation for various gas pressures and we
plot $\varphi(n_{gas})$ and $-\ln[t(n_{gas})]$ as a function of
pressure (see Fig. \ref{fig2}). These two quantities are expected
to vary linearly with the gas density
\cite{schmiedmayer95,schmiedmayer97,hammond97} and the high signal
to noise ratio of our experiments allows us to confirm these
theoretical expectations. To deduce $(n-1)$ from this plot, we
need the $k L$ value. $k$ is calculated from the lithium beam mean
velocity $u$ measured by Bragg diffraction. The effective cell
length $L$ is calculated by weighting each element $dz$ by the
local gas density. In the molecular regime, the density in the
slits varies linearly with $z$ and vanishes near the slit exit
\cite{beijerinck75}. The effective length $L$ is then the sum of
the inner part length and of the mean of the slit lengths, $L =
66.5 \pm 1.0$ mm. Our final results are the real and imaginary
parts of $(n-1)$ divided by the gas density and the dimensionless
ratio $\rho = {\mathcal{R}}e(n-1) /{\mathcal{I}}m(n-1)$. These
results are collected in table \ref{index1065} for a lithium beam
mean velocity $u = 1075 \pm 20$ m/s and we have similar data for
several other velocities.

\begin{figure}
\includegraphics[width = 7.5 cm,height= 5.6 cm]{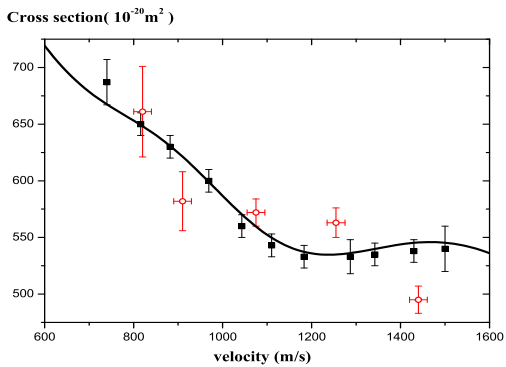}
\caption{Plot of the lithium-xenon total collision cross section
$\left<\sigma\right>$ as a function of the lithium mean velocity
$u$ in m/s:  our measurements by atom interferometry (open
circles), measurements of Dehmer and Wharton (squares) and
calculated values (full curve) obtained with the Buckingham-Corner
potential fitted by these authors \cite{dehmer72}.} \label{fig3}
\end{figure}

The imaginary part of the index of refraction, which measures the
attenuation of the atomic beam by the gas, is related to the total
collision cross section $\left<\sigma\right>$ by
$\left<\sigma\right>  = 2 {\mathcal{I}}m(n-1)k/n_{gas}$ where
$\left<\right>$ designates the average over the target gas thermal
velocity. Fig. \ref{fig3} compares the cross section
$\left<\sigma\right>$ deduced from our index measurements with the
values obtained by L. Wharton and co-workers by scattering
techniques \cite{ury72,dehmer72}: the agreement is very good,
although the velocity distribution of our lithium beam, with a
full width at half maximum close to $25$\%, is broader than the
$4.4$\% FWHM distribution used by L. Wharton and co-workers.

\begin{figure}
\includegraphics[width = 7.5 cm,height= 5.6 cm]{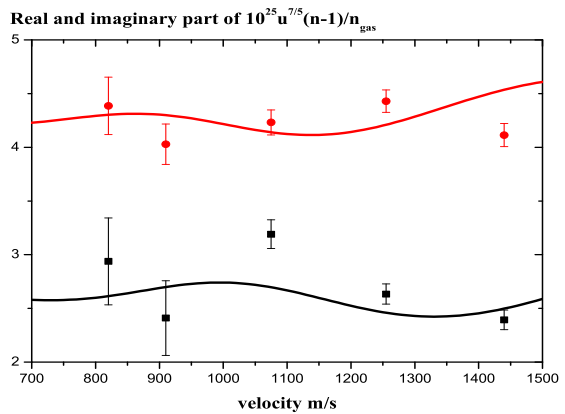}
\caption{Plot of the real (squares) and imaginary (dots) parts of
$10^{25} u^{7/5}(n-1) /n_{gas}$ measured for xenon as a function
of the lithium beam mean velocity $u$ ($n_{gas}$ in m$^{-3}$, $u$
in m/s). The full curves represent the calculated values
\cite{champenois97}, using the Buckingham-Corner potential of ref.
\cite{dehmer72}. The agreement between the measured and calculated
values is good, especially as there are no free parameters.
}\label{fig4}
\end{figure}

From theory \cite{champenois97,champenois99}, we know that $(n-1)$
decrease with the lithium velocity $u$, like $u^{-7/5}$, with
glory oscillations superimposed on this variation. We suppress
this rapid variation by plotting the real and imaginary parts of
$u^{7/5}(n-1)/n_{gas}$: Fig. \ref{fig4} presents such a plot in
the case of xenon, with our measurements and calculated values
obtained by C. Champenois in her thesis \cite{champenois97} using
the Buckingham-Corner potential fitted by Dehmer and Wharton
\cite{dehmer72}.

\begin{figure}
\includegraphics[width = 8 cm,height= 6 cm]{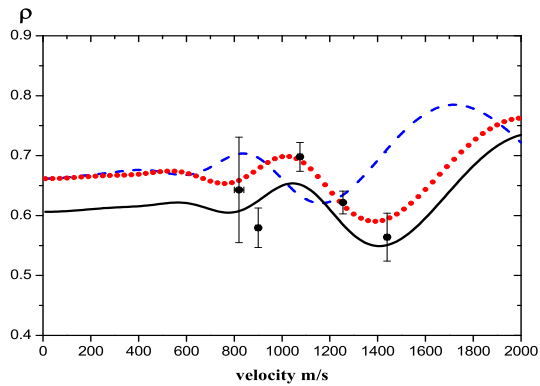}
\caption{ Plot of the  ratio $\rho = {\mathcal{R}}e(n-1)
/{\mathcal{I}}m(n-1)$ for xenon as a function of the lithium beam
mean velocity $u$. The points are our experimental values while
the curves have been calculated with the lithium-xenon potential
of references  \cite{dehmer72} (full line),  \cite{cvetko94}
(dotted line) and  \cite{patil91} (dashed line).}\label{fig5}
\end{figure}

Figure \ref{fig5} compares our measurements of the ratio $\rho=
{\mathcal{R}}e(n-1) /{\mathcal{I}}m(n-1)$ with calculations. We
have chosen the case of xenon for which we have more data points.
The mean $\rho$ value is lower than the $\rho = 0.726$ value
predicted by the group of D. Pritchard
\cite{schmiedmayer95,schmiedmayer97} for a purely attractive
$r^{-6}$ potential. A lower mean $\rho$ value is expected when the
$n=8,10$ terms of the $r^{-n}$ expansion of the long range
potential are also attractive and not negligible
\cite{champenois97}. Moreover, a glory oscillation is clearly
visible on our measurements as well as on the calculations done by
C. Champenois \cite{champenois99} with three different potential
curves: two ab initio potentials \cite{cvetko94,patil91} and the
Buckingham-Corner potential of Dehmer and Wharton \cite{dehmer72}.
The three calculated curves reproduce well the observed amplitude
of the glory oscillation. The observed phase is close to the
calculated phase for the potentials of references
\cite{cvetko94,dehmer72} and not for the one of reference
\cite{patil91}. This is not surprising as this phase is very
sensitive to the potential well depth \cite{champenois97}.

In this letter, we have described the first measurements of the
index of refraction of gases for lithium waves, with an experiment
similar to those performed by D. Pritchard and co-workers with
sodium waves \cite{schmiedmayer95,schmiedmayer97,roberts02}. A gas
cell, introduced on one of the atomic beams inside an atom
interferometer, modifies the wave propagation and this
modification is detected on the interference signals. We have
measured the real and imaginary parts of the index of refraction
$n$ for three gases and several lithium velocities and we have
verified the linear dependance of $(n-1)$ with the gas density
$n_{gas}$. Our measurements of the imaginary part of $(n-1)$ are
in very good agreement with previous measurements of the total
cross-section and the real part, which can be measured only by
atom interferometry, is also in good agreement with calculations
of this quantity. Moreover, the comparison between experimental
and theoretical values of the ratio $\rho =
{\mathcal{R}}e(n-1)/{\mathcal{I}}m(n-1)$ is already able to favor
certain interaction potentials.

We hope to improve our experiment, in particular the septum which
limits our ability to use higher velocities of the lithium beam.
The measurement of the index of refraction in a larger velocity
range and with an improved accuracy should then provide a
stringent test of interaction potentials. However, it seems clear
that index of refraction measurements cannot be inverted to
provide a very accurate interaction potentials, as long as the
target gas is at room temperature, because thermal averaging
washes out the low-energy resonances and most of the glory
oscillations. An experiment with a target gas at a very low
temperature should provide a lot more information
\cite{forrey97,blanchard03,audouard95}. Finally, if the target gas
and the atomic beam were both ultra-cold, one could study the
index of refraction in the quantum threshold regime and measure
the atom-atom scattering length as well as low energy resonances
\cite{kharchenko01,vigue95}. The collision experiment involving
two ultra-cold atom clouds, which was made by J. Walraven and
co-workers \cite{buggle04}, proves the feasibility of dense enough
gas targets for collision studies. The development of a cold atom
interferometer coupled to such a gas target remains to be done.

Finally, the interaction of a matter wave with a gas induces
decoherence. This effect was studied by Hornberger et al. with
C$_{70}$ in a Talbot Lau interferometer \cite{hornberger03} and by
Uys et al. with sodium in a Mach-Zehnder interferometer
\cite{uys05}: in both cases, a low pressure of gas is introduced
everywhere inside the interferometer. This decoherence process
depends on the momentum transferred to a particle of the wave by a
collision with an atom of the scattering gas and on the
interferometer geometry (separation between the two arms, size of
the detector). This decoherence effect is thus quite different
from the index of refraction for which all scattering events
contribute.

\acknowledgments

We have received the support of CNRS MIPPU, of ANR and of R\'egion
Midi Pyr\'en\'ees. The technical staff, G. Bailly, D. Castex, M.
Gianesin, P. Paquier, L. Polizzi, T. Ravel and W. Volondat, has
made these experiments possible. We thank former members of the
group A. Miffre, C. Champenois and N. F\'elix, for their help and
A. Cronin, for fruitful advice.

\end{document}